\documentstyle[pre,multicol,aps]{revtex}

\begin{document}
\input{epsf.tex}

\title{Multisoliton complexes in a sea of radiation modes}

\author{Andrey A. Sukhorukov, Adrian Ankiewicz, and Nail N. Akhmediev}
\address{
Optical Sciences Centre, Research School of Physical Sciences and Engineering, Australian National University,\\ 
Canberra, ACT 0200, Australia
}
\maketitle

\begin{abstract}
We derive exact analytical solutions describing multi-soliton complexes and their interactions on top of a multi-component background in media with self-focusing or self-defocusing Kerr-like nonlinearities. These results are illustrated by numerical examples which demonstrate soliton collisions and field decomposition between localized and radiation modes.
\end{abstract}

\vspace*{-1.0 cm}

\begin{multicols}{2}
\narrowtext

\section{Introduction}

The multi-soliton complex (MSC) is a new concept of modern nonlinear science~\cite{biondini}. If we think of solitons as elementary particles, then we can imagine multi-soliton complexes (MSCs) as combinations of elementary particles such as atoms. In most cases an atom is an isolated particle, but it can demonstrate more complicated behavior. Mathematically speaking, a multi-soliton complex is a self-localized state which is a nonlinear superposition of several fundamental solitons. As it is a composite structure, we can expect it to behave in a more complicated way than a single fundamental soliton. At the same time, an MSC can behave as a single particle unless there is a substantial external perturbation which causes it to split into its primary constituents.

There are various phenomena in nonlinear physics which can be treated as MSCs, some of which are listed in Ref.~\cite{biondini}. Here, we are mainly interested in optical applications. In fiber optics, an MSC is a single short pulse formed by a nonlinear superposition of fundamental solitons trapped in a common potential well, so that all of them have the same velocity~\cite{Hasegawa2}. The nonlinear superposition can be either coherent or incoherent, in the sense that the phases of the separate solitons in the bunch can be either related or independent. The solitons in the group may be bound together, or at least may stay close to each other, simply because they have the same initial speed. The latter happens in the case of integrable systems.

Incoherent solitons in space~\cite{Christod1,Christod2} have recently attracted considerable attention~\cite{PT}, especially after the first experimental demonstration of their existence~\cite{Incohfirst}. Incoherent self-trapping in a biased photorefractive crystal is usually well-described by a set of $M$ coupled nonlinear Schr\"odinger equations (NLSEs) with saturable nonlinearities~\cite{Christod2}. Moreover, in photorefractive media with a drift mechanism of nonlinear response, the Kerr-like nonlinearity is a good model~\cite{Vysl2}. In the latter case, the equations become integrable, and this allows us to obtain solutions analytically. Moreover, the solution can be considered as a nonlinear superposition of a certain number of fundamental solitons and radiative waves. This fact allows a simple qualitative approach to the problem, in addition to having an exact solution.

The number of solitons in a complex depends on the nature of physical system, as well as on the initial conditions. Mathematically speaking, we will talk about $N$ fundamental solitons all having the same speed and moving (or resting) as a single complex. The value of $N$ can be~$1$ (fundamental soliton) but can take large values up to infinity.

Elementary particles in nature never exist in isolation, but are submerged in a fluctuating radiation field. Similarly, MSCs can exist on top of a background plane wave and, more generally, on top of an arbitrary number of plane waves propagating in different directions. In this sense, we will talk about {\em a sea of radiation modes}, which can be stable with respect to modulational perturbations in both self-focusing and defocusing media~\cite{mod_stab}. Plane waves can also serve as a basis for the existence of dark solitons. An incoherent soliton is an example of an MSC, and stable propagation of both dark and bright spatial optical solitons on an incoherent background has been observed in recent experiments~\cite{incoh_backgr_df,incoh_backgr_sf}.

In this work we present a class of exact solutions to the coupled set of $M$ equations which describe symmetric and asymmetric MSCs on a background which is formed from a nonlinear superposition of radiation modes. This differs from the earlier results presented in Refs.~\cite{PRL99,backgr} where the background consisted of a single mode, and Refs.~\cite{incoh_backgr_df,incoh_backgr_sf}, where only some stationary solutions with ${sech^2}$-type intensity profiles were considered. Our new exact solution of MSCs in a sea of radiation modes accounts for an arbitrary number of both soliton and radiative components.

\section{Model equations and their properties}

It can be shown~\cite{Vysl2} that, for photorefractive media with a drift mechanism of nonlinear response, a good approximate model describing the propagation of $M$ self-trapped mutually incoherent wave packets in a planar waveguide is the set of NLSEs for a Kerr-type nonlinearity
\begin{equation} \label{eq:model}
  i \frac{\partial \psi_{m}}{\partial z}
  + \frac{1}{2} \frac{\partial^{2}\psi_{m}}{\partial x^{2}}
  + \alpha \delta n \, \psi_{m}
  = 0 ,
\end{equation}
where $\psi_m$ denotes the $m$-th component of the beam, $x$ is the transverse coordinate, $z$ is the coordinate along the direction of propagation, $\alpha$ is the coefficient representing the strength of nonlinearity, and $\delta n$ is the change in refractive index profile created by all the incoherent components in the light beam. Because the response time of the nonlinearity is assumed to be long compared with temporal variations of the relative phases of all the components, the medium responds to the average light intensity, and this is just a simple sum of modal intensities~\cite{snyder}. Then, after defining a new set of functions $u_m(x,z) = \sqrt{|\alpha|} \psi_m(x,z)$, we reduce Eq.~(\ref{eq:model}) to the following set of normalized equations:
\begin{equation} \label{eq:nlse}
  i \frac{\partial u_{m}}{\partial z}
  + \frac{1}{2} \frac{\partial^{2}u_{m}}{\partial x^{2}}
  + s u_m \sum_{j=1}^M |u_j|^2
  = 0 .
\end{equation}
Here $1 \le m \le M$, $M$ is the number of components, while $s=+1$ in a self-focusing medium and $s = -1$ in a defocusing material.

Let us briefly outline some general properties of Eqs.~(\ref{eq:nlse}). The most remarkable fact is that these coupled equations are {\em completely integrable} by means of the inverse scattering technique (IST)~\cite{Manakov,ZS,Gerd}. Specifically, any solution can be represented as a nonlinear superposition of {\em solitary waves} and {\em radiation modes} which correspond to the discrete and continuous parts of the IST spectrum, respectively.

Every {\em fundamental soliton} (labeled~$j$) is characterized by several eigenvalues: (i)~a complex wavenumber $k_j = r_j + i \mu_j$, (ii)~a shift in the coordinate plane $(x_j,z_j)$, and (iii)~a polarization vector $\mathbf{p}^{(j)}$ in the functional space, normalized to unity as $\sum_{m=1}^M \left| p_m^{(j)} \right|^2 = 1$. The simplest bright single--soliton solution in a self-focusing medium ($s=+1$) can be written as:
\begin{equation} \label{eq:one_soliton}
  u_m(x,z) = p^{(j)}_m \; r_j \; {\rm sech}( \beta_j ) \; e^{i \gamma_j}  ,
\end{equation}
where $\beta_j = r_j (\bar{x}_j - \mu_j \bar{z}_j)$, $\gamma_j = \mu_j \bar{x}_j + (r_j^2-\mu_j^2) \bar{z}_j / 2$, and $(\bar{x}_j,\bar{z}_j) = (x - x_j,z - z_j)$ are the shifted coordinates. We see that the peak soliton intensity and its inverse width are determined by the real part of the wavenumber $r_j$, while the imaginary part $\mu_j$ characterizes the soliton velocity in the transverse direction. Moreover, each fundamental soliton can be ``spread out'' into several incoherent components, as defined by the polarization vector.

The solution for a single radiation mode, in the form of a plane wave, can also be characterized by a similar set of parameters,
\begin{equation} \label{eq:one_rmode}
  u_m(x,z) = p^{(j)}_m \; r_j \; e^{i \sigma_j } ,
\end{equation}
where $\sigma_j = \mu \bar{x}_j + (2 s r_j^2 - \mu_j^2) \bar{z}_j / 2$. Such a plane wave exists for either sign of nonlinearity, $s = \pm 1$, and it is stable in a self-defocusing medium. Moreover, an incoherent superposition of a large number of plane waves can be stable even in the self-focusing medium, as was shown in~\cite{mod_stab}. In the presence of solitons, the plane waves are distorted, but due to the integrability of the original equations, the corresponding solutions can be obtained in an explicit form, as we demonstrate in the following section.

\section{Explicit solutions for multi-soliton complexes on a background}

It has been demonstrated in earlier studies~\cite{backgr,mprl} that {\em a stationary MSC} can only be formed by {\em incoherently coupled} fundamental solitons and radiation waves which have orthogonal polarization vectors, i.e. $\sum_{m=1}^M p_m^{(j)} p_m^{(n)} = \delta_{jn}$. We restrict our investigation to this important case. Then, the mathematical description can be simplified if we use the {\em rotational symmetry of the functional space} of the original Eqs.~(\ref{eq:nlse}). Indeed, it is sufficient to find solutions $u_j$ where all the fundamental nonlinear eigenmodes belong to different components, $p_m^{(j)} = \delta_{mj}$, and then the full family of solutions can be determined using the following transformation:
\begin{equation} \label{eq:rotate}
   \bar{u}_m = \sum_{j=1}^M R_{mj} u_j ,
\end{equation}
where the matrix $R_{mj}$ defines a rotation in the $M$-dimensional space (characterized by \mbox{$M-1$} angles), which preserves the MSC intensity profile $\sum_m |\bar{u}_m|^2 \equiv \sum_m |{u}_m|^2$. 

As follows from the IST, it is possible to construct the full solution by adding the radiation modes to bright MSCs. On the other hand, solutions for bright MSCs, having a zero far-field asymptotic in the self-focusing medium ($s=+1$) can be found by solving a set of linear equations~\cite{Nogami,Gardner}:
\begin{equation} \label{eq:slau}
  \sum_{m=1}^{M_s} \frac{e_j e_m^{\ast} u_m}{k_j +k_m^{\ast}}
         + \frac{1}{2 r_j} u_j = - e_j ,
\end{equation}
where $M_s$ is the number of fundamental solitons, and $e_j = \chi_j \exp \left( \beta_j + i \gamma_j \right)$. Here the coefficients $\chi_j$ define the relative coordinate centers for the fundamental solitons. By choosing their values in a special way, the solutions of Eqs.~(\ref{eq:slau}) can be presented in a simple, symmetric form (see Refs.~\cite{mprl,slv4} for details).

In order to reveal the basic properties of radiation modes under the presence of an MSC, we perform {\em a linear analysis}, assuming that the wave amplitudes are vanishingly small and that they do not contribute to the intensity profile. We note that Eqs.~(\ref{eq:slau}) define an extension of the functional basis introduced by Kay and Moses~\cite{Kay} to construct reflectionless potentials. The difference is that in Ref.~\cite{Kay} the $k_j$ are real, whereas we performed a similar analysis and found that the results can be generalized to the case of complex wavenumbers. Specifically, we derived solutions for scattered plane waves, or radiation modes belonging to the continuous spectrum:
\begin{equation} \label{eq:scatter}
   u_m = r_m \left( 1 + \sum_{j=1}^{M_s}
                    \frac{ u_j e_j^{\ast} }{ k_j^{\ast} + i \mu_m} \right)
              e^{i \mu_m (\bar{x}_m - \mu_m  \bar{z}_m / 2) } .
\end{equation}
In general, we can have an arbitrary number ($M_r$) of radiation modes, and we assume that they belong to the components with $M_s+1 \le m \le M_s + M_r$. Note that solution (\ref{eq:scatter}), which is valid in the limit $r_m \rightarrow 0$, reduces to a simple plane-wave profile given by Eq.~(\ref{eq:one_rmode}) in the absence of bright components. When the radiation wave amplitudes $r_m$ are not small, both the radiation modes and the bright soliton profiles defined by Eq.~(\ref{eq:slau}) are distorted according to the nonlinear superposition principle. In order to find MSC solutions on a finite background, we extend the approach introduced in Ref.~\cite{backgr}. First, we multiply Eqs.~(\ref{eq:slau}) by $u_j^{\ast} / (k_j - i \mu)$, add the complex conjugate, and sum over the fundamental soliton numbers $1 \le j \le M_s$. Comparing the resulting expression with Eq.~(\ref{eq:scatter}), we obtain the following relation:
\begin{equation} \label{eq:rot_one}
  \sum_{j=1}^{M_s} \left| \frac{u_j}{k_j^{\ast} + i \mu_m} \right|^2
  + \left|\frac{u_m}{r_m} \right|^2 = 1 ,
\end{equation}
where the subscript $m$ indicates a radiation mode. The second step is to define the {\em combined rotational--scaling transformation} 
which changes the nonlinear eigenmode amplitudes while preserving the total intensity profile up to a constant background,
\begin{equation} \label{eq:I_shift}
  \tilde{I} = I_b + s I .
\end{equation}
Here $\tilde{I}$ is the full intensity of the re-scaled solution $\tilde{u}_m (x,z)$, and $I_b = \sum_{j=M_s+1}^{M_s+M_r} |r_j|^2$ is the background intensity. Then, from Eq.~(\ref{eq:rot_one}), it immediately follows that the rescaling coefficients for the bright components should be defined as 
\begin{equation} \label{eq:U_m}
  |U_m|^2 = s + \sum_{j=M_s+1}^{M_s+M_r}
                   \frac{r_j^2}{|k_m^{\ast} + i \mu_j|^2},
\end{equation}
while for the radiation modes we put $U_m = 1$. Quite remarkably, this procedure can be used for both signs of nonlinearity ($s = \pm 1$). The propagation constants should also be modified accordingly, so that the resulting functions satisfy the original Eqs.~(\ref{eq:nlse}). We finally obtain:
\begin{equation} \label{eq:slv_rot}
   \tilde{u}_m (x,z) = U_m e^{i s I_b z } u_m(x,z) .
\end{equation}
At this point, the derivation of the analytical solutions for the MSCs existing on top of several radiation modes is complete, and the component profiles are defined by Eqs.~(\ref{eq:U_m}) and~(\ref{eq:slv_rot}) together with Eqs.~(\ref{eq:slau}) and~(\ref{eq:scatter}). All such solutions correspond to multi-parameter families which can be generated with the help of the rotation transformation~(\ref{eq:rotate}).

Let us now extend the analytical results to the case where the background is composed of a continuum set of radiation modes, i.e. $M_r \rightarrow +\infty$. This case corresponds, for example, to spatial optical solitons excited by an incoherent light source~\cite{incoh_backgr_df,incoh_backgr_sf}. Then, the plane waves in the background can be characterized by an angular distribution function, $R(\mu)$ ($\ge 0$), so that $R(\mu) d \mu$ is the wave intensity corresponding to the tangents of inclination angles in the interval $(\mu, \mu + d \mu)$. Therefore, the full background intensity is $I_b = \int_{-\infty}^{+\infty} R(\mu) d \mu$, and the scaling coefficients for bright components are
\begin{equation} \label{eq:U_m_int}
   |U_m|^2 = s + \int_{-\infty}^{+\infty}
                     \frac{R(\mu)}{|k_m^{\ast} + i \mu|^2} d \mu .
\end{equation}
We note that, for a finite number of radiation modes, the distribution function can be written as $R(\mu) = \sum_{j=M_s+1}^{M_s+M_r} r_j^2 \delta(\mu-\mu_j)$, and then expression~(\ref{eq:U_m_int}) reduces to Eq.~(\ref{eq:U_m}).

The above results are valid for both self-focusing ($s=+1$) and self-defocusing ($s=-1$) media. As follows from Eq.~(\ref{eq:I_shift}), the qualitative difference is that, in the former case we have bright complexes on a constant background while in the latter case dark dips are formed.

\section{Interaction properties of MSC\lowercase{s}}
                   \label{sect:interaction}

The general analytical solutions can be used to obtain some important characteristics in a simple form. In particular, we can use the fact that the outcome of a soliton collision, in terms of total intensity, is not affected by the presence of the background. Therefore, according to earlier results~\cite{mprl}, the interactions of MSCs are characterized by translational shifts of fundamental solitons after the collision, and these are given by
\begin{equation} \label{eq:shift}
   \delta x_j = \frac{1}{r_j} \sum_m f_{jm}
           \log\left| \frac{k_j + k_m^{\ast}}{k_j - k_m} \right|.
\end{equation}
Here the summation involves the fundamental solitons which feature in the collisions with the soliton number $j$. When the colliding soliton number $m$ comes from the right (i.e. has a larger $x$ coordinate before the impact), we put $f_{jm} = +1$ while we set $f_{jm} = -1$ when it comes from the left. The shifts differ for each soliton in an MSC. As a result, the
intensity profile of an MSC changes after collisions.

\section{Modulation of background components} \label{sect:modulation}

According to the general relation~(\ref{eq:I_shift}), the intensity profile is uniquely determined by the eigenvalues of the bright fundamental solitons and the background intensity $I_b$, but does not depend on the the angular distribution of radiation waves. However, the total intensity of the soliton components,
\[
    I_s = \sum_{m=1}^{M_s} |U_m|^2 |u_m|^2,
\]
and the intensity of the radiation modes,
\begin{equation} \label{eq:Ir}
   I_r = I_b - \sum_{m=1}^{M_s} (|U_m|^2-s) |u_m|^2,
\end{equation}
both depend on the scaling coefficients $U_m$ defined in Eq.~(\ref{eq:U_m_int}). As follows from Eq.~(\ref{eq:Ir}), each fundamental soliton creates a hole in the background, and the corresponding modulation depth is proportional to the {\em bright-dark coupling coefficient}, given by the value $(|U_m|^2 - s)$. Quite interestingly, the radiation mode profiles are the same in self-focusing and self-defocusing media, provided the distribution function and soliton eigenvalues remain unchanged.

However, there are some key differences between solitons in self-focusing and self-defocusing media. In the former case, a modulation of the background is compensated by the bright components having larger amplitudes (since $|U_m|^2 > s$, and $s=+1$). On the other hand, in a self-defocusing medium, a dark soliton creates an effective waveguide, which in turn can trap bright components. Such a self-trapping mechanism results in {\em the limitation of the minimum dark soliton width}. This happens because the maximum intensity contrast is limited by the value of the background intensity. As a matter of fact, the limitation can be even stricter, since the maximum modulation depth, ${\cal M} = {\rm max}_x (I_b-I) / I_b < 1$, cannot always reach the value of $1$. Then, according to Eq.~(\ref{eq:one_soliton}), the characteristic width corresponding to one fundamental soliton cannot exceed the value $({\cal M} I_b)^{-1/2}$. The actual limit is determined by solving the existence conditions, which follow from the requirement for the right-hand side of Eq.~(\ref{eq:U_m_int}) to be non-negative, since, by definition, $|U_m|^2 \ge 0$. It is interesting to note that these conditions involve only the individual wavenumbers of fundamental solitons, and they are automatically satisfied for interacting solitons forming MSCs.

Finally, we note that the radiation modes are characterized by a non-trivial phase modulation. For practical applications, it is especially important to know the {\em phase jump}, or the additional phase shift which appears due to the presence of bright fundamental solitons. Using Eqs.~(\ref{eq:slau}) and~(\ref{eq:scatter}), we find the following relation,
\[
     e^{i \phi(\mu)}
        = \prod_{m=1}^{M_s} \frac{i \mu - k_m}{i \mu + k_m^{\ast}} ,
\]
where $\phi$ is the phase jump, and $\mu$ defines the inclination angle of the radiation mode. Then, the phase jump can be found as a sum over the phase shifts associated with individual fundamental solitons,
\begin{equation} \label{eq:phase}
   \phi(\mu) = \sum_{m=1}^{M_s} \phi_m(\mu)
             = \sum_{m=1}^{M_s} 2
                      \arctan \left( \frac{r_m}{\mu - \mu_m} \right) .
\end{equation}
We see that the absolute values of the individual phase shifts are limited to $\pi$. However, the total phase jump can become larger than $\pi$ if $M>1$, i.e. if the MSC is composed of several fundamental solitons.

\begin{figure}
\setlength{\epsfxsize}{6cm}
\centerline{\mbox{\epsffile{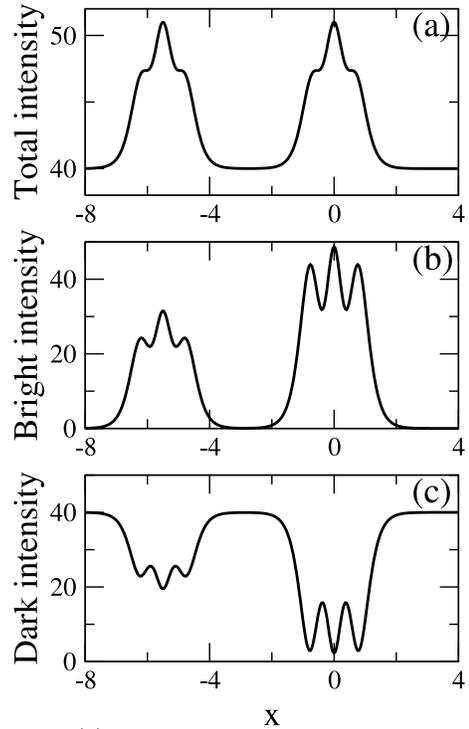}}}
\caption{\label{fig:sf1}
(a)~Total intensity profiles for multisoliton complexes, and the corresponding decomposition between (b)~bright and (c)~dark components in a self-focusing medium.
Each MSC consists of three fundamental solitons with $r_m=2,3,4$; the MSC on the right is stationary ($\mu_m=0$), while the MSC on the left has a positive velocity ($\mu_m = 3$). Angular width of background distribution is $\rho=1$.
}
\end{figure}

\section{General results for a Gaussian angular distribution}
                                   \label{sect:gauss}

Let us now analyse the features of the bright-dark decomposition for a Gaussian-type angular distribution,
\begin{equation} \label{eq:Gauss}
R(\mu) = I_b \frac{1}{\rho \sqrt{\pi}} e^{-\mu^2 / \rho^2} ,
\end{equation}
where $\rho>0$ is the characteristic angular width. Since the integral in Eq.~(\ref{eq:U_m_int}) cannot be expressed in elementary functions for arbitrary $\rho$, we first consider the limiting cases. Specifically, for {\em a narrow angular distribution}, i.e. $\rho \ll 1$, we have $(|U_m|^2-s) \simeq I_b / (r_m^2+\mu_m^2)$, while for $\rho \gg 1$ (and $\rho \gg \mu_m$) we obtain $(|U_m|^2-s) \simeq I_b \sqrt{\pi} / (\rho r_m) \rightarrow 0$. Therefore, we expect that, for a fixed background intensity $I_b$, the modulation of the radiation waves should be reduced (i)~for wider angular distributions, i.e. larger $\rho$, and (ii)~for MSCs having higher velocities $|\mu_m|$.

On the other hand, since the phase jump depends on the radiation mode wave number $\mu$, the excitation of solitons can be more difficult in cases of wider angular spectra of radiation modes, i.e. larger $\rho$. Additionally, for a moving MSC, i.e. when $\mu_m = {\rm const} \ne 0$, the dependence $\phi(\mu)$ becomes asymmetric unless $R(\mu-\mu_m) = R(\mu_m-\mu)$.

\begin{figure}
\setlength{\epsfxsize}{6cm}
\centerline{\mbox{\epsffile{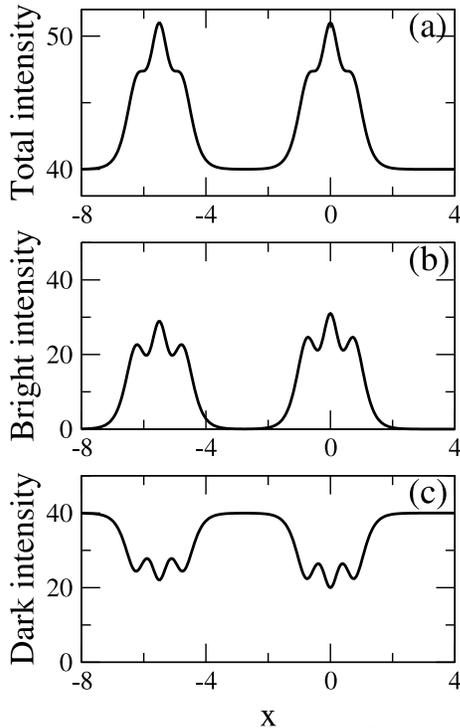}}}
\caption{\label{fig:sf2}
Intensity distributions for $\rho=7$ in a self-focusing medium. Parameters and notation are the same as in Fig.~\ref{fig:sf1}.
}
\end{figure}

\section{Bright MSC in a self-focusing medium} \label{sect:bright}

For a self-focusing nonlinearity ($s=+1$), MSCs exist in the form of bright localized waves having higher intensity than the background, as follows from Eq.~(\ref{eq:I_shift}). Examples of the total intensity profiles and the bright-dark mode decomposition are shown in Figs.~\ref{fig:sf1} and~\ref{fig:sf2} for different values of $\rho$. The MSC on the right has zero velocity (with the corresponding $\mu_m=0$), while the other MSC (on the left) has a positive velocity; thus in the latter case, the corresponding background modulation is smaller, as predicted in Sec.~\ref{sect:gauss}.

\begin{figure}
\vspace*{-8mm}
\setlength{\epsfxsize}{8cm}
\centerline{\mbox{\epsffile{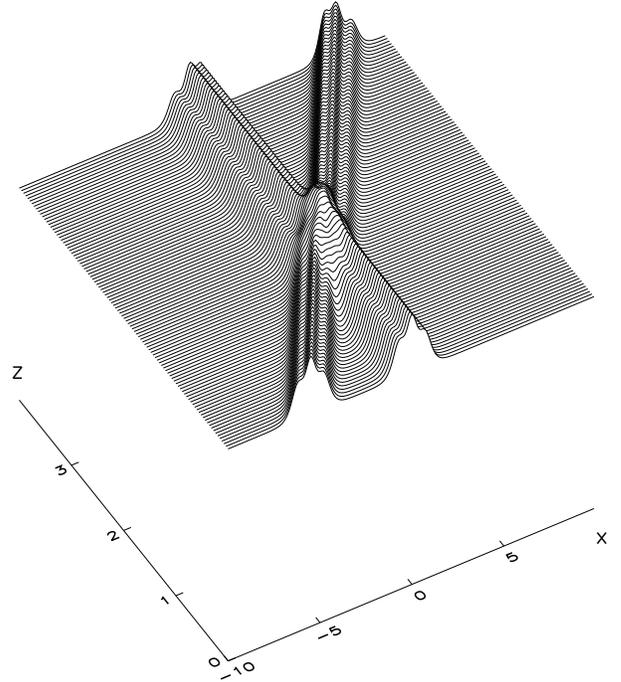}}}
\vspace*{0.5cm}
\caption{\label{fig:sf_collis}
Interaction of two multisoliton complexes existing on a multi-component background in a self-focusing medium. Input profile corresponds to Fig.~\ref{fig:sf1}.
}
\end{figure}

A collision between two MSCs is illustrated in Fig.~\ref{fig:sf_collis}. This example corresponds to the initial conditions shown in Fig.~\ref{fig:sf1}. A remarkable fact is that the total intensity profile does not depend on the value of $\rho$, provided that $I_b$ is preserved. The intensity profile for the collision will be the same for other values of $\rho$ or for other distribution functions. In these examples, the MSC actually has an intensity which is relatively small compared with the background level.

Note that the shape of each MSC changes after a collision, for the reasons discussed in Sec.~\ref{sect:interaction}. In particular, a symmetric MSC becomes asymmetric after a collision (see also Ref.~\cite{mprl}]). The presence of radiation does not influence this process. Another feature of a collision is that the lateral shift of the MSCs is relatively large. For example, it can easily be seen on the scale of Fig.~\ref{fig:sf_collis}. In contrast to single solitons, MSCs experience larger shifts in collisions, due to the multiple contributions from all the constituent fundamental solitons.

\section{Dark solitons in self-defocusing medium} \label{sect:dark}

\begin{figure}
\setlength{\epsfxsize}{8cm}
\centerline{\mbox{\epsffile{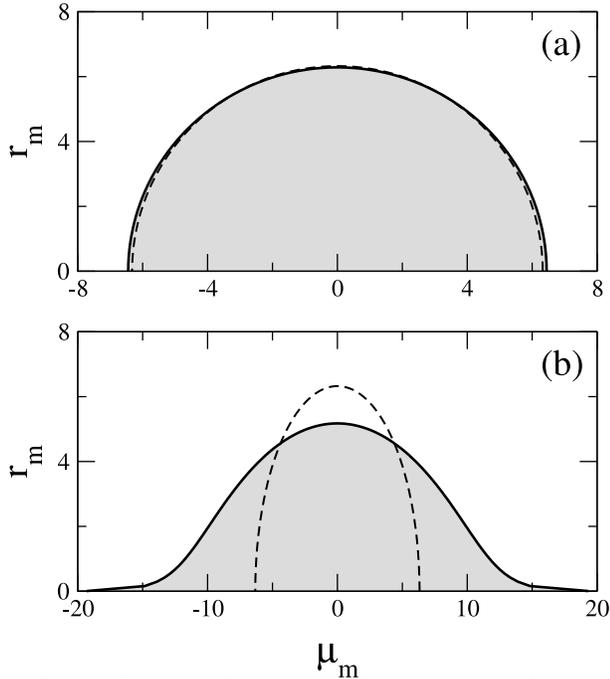}}}
\caption{\label{fig:df_exist}
Grey shading marks existence regions in the parameter space of fundamental soliton eigenvalues $(r_m,\mu_m)$. The angular distribution function is given by Eq.~(\ref{eq:Gauss}) with $I_b=40$ and (a)~$\rho=1$ or (b)~$\rho=7$. Dashed lines correspond to the case of a single component background, when $\rho \rightarrow 0$.
}
\end{figure}

To describe MSCs on a background in a self-defocusing medium ($s=-1$), we first have to determine the existence conditions, as outlined in Sec.~\ref{sect:modulation}. Considering the case of the Gaussian distribution given by Eq.~(\ref{eq:Gauss}), we find that, for a narrow angular spectrum, in the lowest-order approximation, the existence condition is $r_m^2+\mu_m^2 \le I_b$. Therefore, the minimum soliton width, which is of order $r_m^{-1}$, can be achieved if the soliton velocity is zero. In the other limit where $\rho \gg 1$ and $\rho \gg \mu_m$, we have $r_m \le I_b \sqrt{\pi} / \rho$, i.e. the minimum width increases linearly with an increase in $\rho$. 
Numerically-calculated existence regions are shown in Fig.~\ref{fig:df_exist} for two values of $\rho$ in Eq.~(\ref{eq:Gauss}). The figure clearly shows that the existence region is very similar to that in the case of a single component radiation field when $\rho$ is relatively small. However, the existence regions become visibly different when $\rho$ is large.

\begin{figure}
\setlength{\epsfxsize}{8cm}
\centerline{\mbox{\epsffile{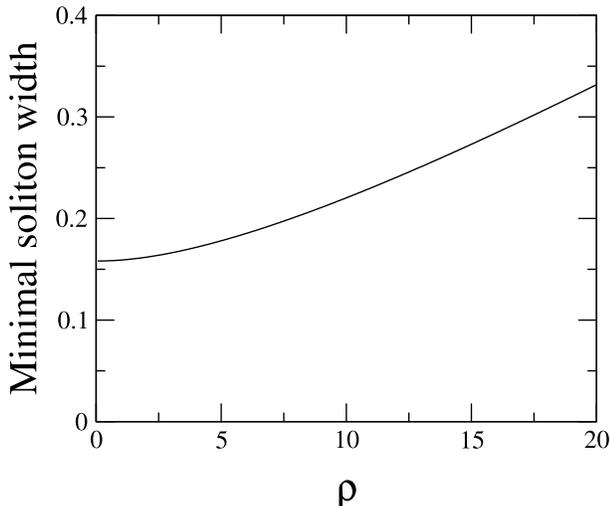}}}
\caption{\label{fig:df_width}
Dependence of the minimum characteristic soliton width on the parameter $\rho$ for the angular distribution function of radiation waves given by Eq.~(\ref{eq:Gauss}) with $I_b=40$.}
\end{figure}

The minimum soliton width versus $\rho$ is shown in Fig.~\ref{fig:df_width}. This result shows that the distribution function for the radiation field influences the properties of an MSC, namely it changes the limiting parameters for the existence of an MSC, although the intensity profile of the MSC is not directly influenced by the properties of the radiation field.

\begin{figure}
\setlength{\epsfxsize}{6cm}
\centerline{\mbox{\epsffile{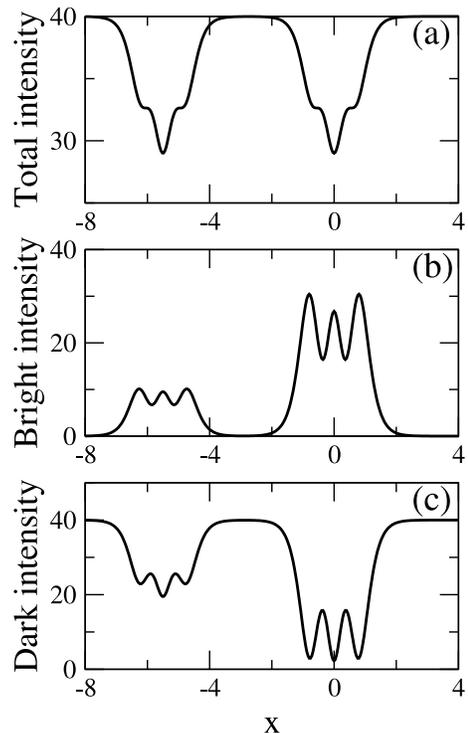}}}
\caption{\label{fig:df1}
Intensity distributions in a self-defocusing medium. Parameters and notation are the same as in Fig.~\ref{fig:sf1}.
}
\end{figure}

Figure~\ref{fig:df1} shows an example of intensity distribution for the case of a self-defocusing medium (i.e. dark MSCs on a background). We have chosen the soliton eigenvalues to be the same as in Figs.~\ref{fig:sf1} and~\ref{fig:sf_collis}. According to our general expression~(\ref{eq:I_shift}), the total intensity profiles in self-defocusing and self-focusing media are ``mirror-images'' relative to the level of the background. Even the radiation mode intensities coincide in these two cases, cf. Figs.~\ref{fig:sf1}(c) and~\ref{fig:df1}(c). However, the bright component intensities are different, as is clearly seen in Fig.~\ref{fig:sf1}(b) and~\ref{fig:df1}(b). This is a manifestation of the nontrivial nature of the nonlinear superposition of the solitons and the background components.

\begin{figure}
\vspace*{-8mm}
\setlength{\epsfxsize}{8cm}
\centerline{\mbox{\epsffile{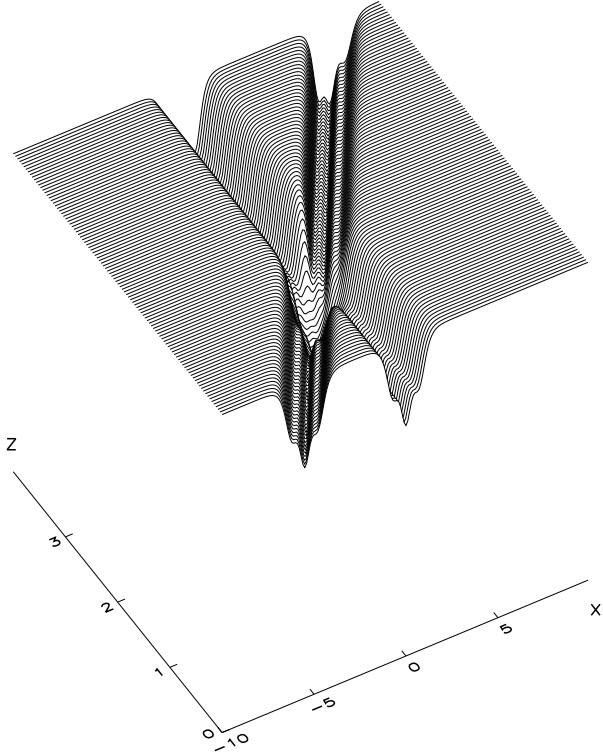}}}
\vspace*{0.5cm}
\caption{\label{fig:df_collis}
Interaction of two multisoliton complexes existing on a multi-component background in a self-defocusing medium. Input profile corresponds to Fig.~\ref{fig:df1}.
}
\end{figure}

Figure~\ref{fig:df_collis} shows the collision of two MSCs on a background. Again, we can see that the nonstationary intensity profile created by the soliton interaction during collision is the "mirror image" of that for bright MSCs in a self-focusing medium, as shown in Fig.~\ref{fig:sf_collis}. The symmetry relation is mathematically exact. Correspondingly, the lateral shift is also governed by the same rules as those for a bright MSC.

An important consequence is that the change of refractive index induced by incoherent MSCs has exactly the same pattern in cases of self-focusing and self-defocusing media with Kerr-type nonlinearity. This remarkable fact can be used to design linear multimode waveguides and X-junctions with special properties. Moreover, the MSCs are controlled by many parameters (fundamental soliton eigenvalues), and therefore a wider variety of requirements can be satisfied in comparison with X-junctions formed by a collision of two solitons.

\section{Conclusions}

In conclusion, we have obtained an exact solution for multisoliton complexes on top of a multi-component background composed of radiation waves in Kerr-type nonlinear media. We have identified similarities and differences between bright and dark MSCs which exist in self-focusing and self-defocusing media, respectively. In particular, we have found that the intensity profiles in these two cases are ``mirror-images'' relative to the level of the background, and that they depend only on the eigenvalues of the fundamental solitons. For example, the reshaping of MSCs after collisions is determined by the lateral shifts of the fundamental solitons and is not affected by background components.
On the other hand, the width of dark solitons has a minimum, and this value depends strongly on the angular distribution of the radiation waves. We have derived analytical estimates of the key soliton characteristics for the case of a Gaussian angular distribution of radiation waves, and presented numerical examples illustrating the principal features of bright and dark MSCs.

\end{multicols}
\end{document}